# Topic Modelling and Event Identification from Twitter Textual Data

Marina Sokolova[1,2,*], Kanyi Huang[1], Stan Matwin[2,3], Joshua Ramisch[1], Vera Sazonova[2], Renee Black[4], Chris Orwa[5], Sidney Ochieng[5], Nanjira Sambuli

[1]University of Ottawa, Ottawa, ON, Canada; {sokolova,Joshua.Ramisch}@uottawa.ca

[2]Institute for Big Data Analytics, Dalhousie University, Halifax, NS, Canada

[3]Institute of Computer Science, Polish Academy of Sciences, Warsaw, Poland; stan@cs.dal.ca

[4]Peacegeeks, Vancouver, BC, Canada; renee@peacegeeks.org

[5]iHUB, Nairobi, Kenya; {chris,sidney}@ihub.co.ke

[*]Corresponding Author:

Marina Sokolova, Institute for Big Data Analytics, Dalhousie University and University of Ottawa, 3232, RGN, 451 Smyth Rd, Ottawa, ON, Canada, K1H 8M5

Email: sokolova@uottawa.ca

## ABSTRACT

The tremendous growth of social media content on the Internet has inspired the development of the text analytics to understand and solve real-life problems. Leveraging statistical topic modelling helps researchers and practitioners in better comprehension of textual content as well as provides useful information for further analysis. Statistical topic modelling becomes especially important when we work with large volumes of dynamic text, e.g., Facebook or Twitter datasets. In this study, we summarize the message content of four data sets of Twitter messages relating to challenging social events in Kenya. We use Latent Dirichlet Allocation (LDA) topic modelling to analyse the content. Our study uses two evaluation measures; Normalized Mutual Information (NMI) and topic coherence analysis, to select the best LDA models. The obtained LDA results show that the tool can be effectively used to extract discussion topics and summarize them for further manual analysis.

## Keywords:

Twitter, statistical topic modelling, model evaluation

# 1. Introduction

In recent years, social media has increasingly come under the microscope in the context of political events in order to enhance understanding of current and emerging issues. While there are a wide variety of areas of focus and implications for such study, topics related to peace and security are of particular interest. Communications and propaganda have played particularly important roles in the most violent episodes of our history, such as during the Rwandan genocide, where radio was used to spread violent, hateful and dehumanizing messages that led to the massacre of more than 800,000 Rwandans, crimes which implicated more than one million perpetrators. Social create both new risks in exacerbating tension, but can also create opportunities for responding to violence and incitement towards the prevention of violence and the escalation, and toward building more peaceful, open and democratic societies.

While practitioners have been looking at more 'small data' examples of how messaging can manifest and with what implications, big data and sentiment analysis tools offer an incredible opportunity to understand commentary and behavior in online spaces, towards better understanding their implications offline. The goal of such study is to help relevant actors, notably governments and society actors, to develop meaningful understanding and responses to these emerging trends towards reducing the risk of violence, and towards using social media in more proactive ways to build trust between conflicting communities.

Kenyans are among the most active and avid users of social media tools in Africa[1] (Facebook is estimated to have over 5 million users and Twitter over 1.7 users as of 2015[2]), meaning that effective use, abuse and regulation of social media can play an important role in affecting the outcome of future conflicts.

We are only at the beginning of understanding the potential for such research. This paper explores what we can learn from specific events in Kenya towards considering how to further develop a broader research agenda moving forward.

# 2. Background

With a large number of people embarking on a trend of actively voicing their opinion online on social networks and forums, social media has become a major source for social data mining (Matwin, 2013). Starting from humble beginnings in 2006-2008, Twitter has grown exponentially from 2009. In 2012 the media daily posted up to 500 million of messages (Figure 1). Emergence of Twitter as a leading social media has

---

[1] http://www.businessdailyafrica.com/Corporate-News/AFRICA-Kenya-is-second-most-active-on-twitter/-/539550/1314290/-/kpxav0/-/index.html

[2] http://www.dotsavvyafrica.com/the-5-biggest-social-media-platforms-in-kenya/

made it a useful resource in development of social-oriented text analysis, as in analysis of commentary disseminated via Twitter on the riots in London and other British cities in August 2011 (Tonkin, Pfeiffer, Tourte, 2012)

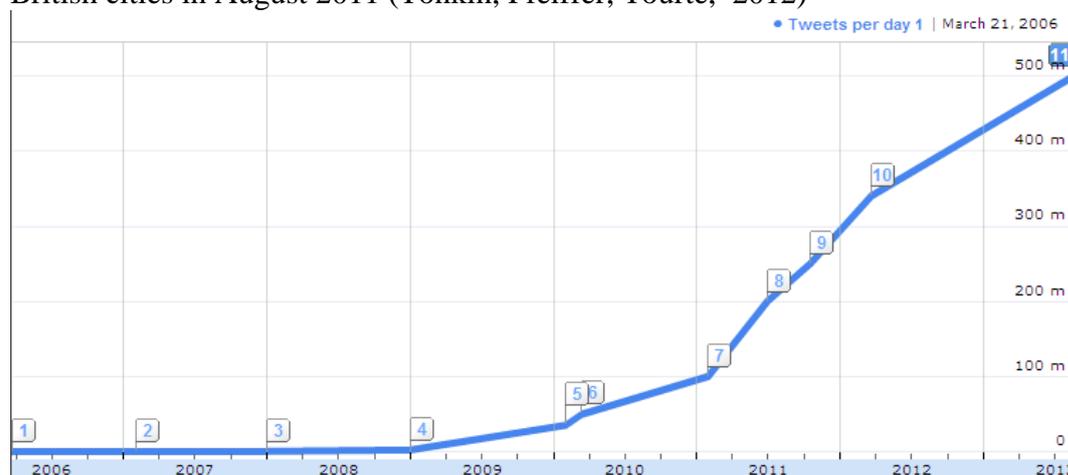

*Figure 1. Growth of posted tweets worldwide per day; m: million; adapted from InternetLiveStats[3]*

Twitter outreach and growth is a worldwide phenomenon, both in the number of Twitter users and posted tweets. According to a survey conducted by *Portland Communications and Tweetminster*, Twitter has become an important source of information in Africa. In 2010 Kenya had the second most active population on Twitter in Africa. At the end of 2013, Nairobi was the most active city in East Africa, with 123,078 geo-located tweets.[4] According to mobile phone operator Airtel Kenya[5], there is constant increase in the number of people following other Kenyan users' accounts (Figure 2). Each follower potentially amplifies the impact of posted messages. This online community, known collectively by the acronym #KOT (Kenyans on Twitter), is now acknowledged as in important and ever-growing force in Kenyan social and political life[6]. With community-based learning prominently contributing to many aspects of Kenyan life (Ramisch et al, 2006), Twitter plays a transformative role in the society. It is also acquiring an important role in both generating – but also curtailing – dangerous speech related to domestic political events, from major events like 2013's general election and the Westgate terror attack, to more localized acts of corruption and political impunity[7].

---

[3] http://www.internetlivestats.com/twitter-statistics/

[4] http://www.portland-communications.com/publications/how-africa-tweets-2014/

[5] http://www.socialbakers.com/resources/client-stories/airtel-kenya/

[6] http://www.nation.co.ke/lifestyle/showbiz/Is-KOT-the--most-powerful-group-in-Kenya-/-/1950810/2515620/-/5omj6y/-/index.html

[7] http://www.bbc.com/news/magazine-33629021

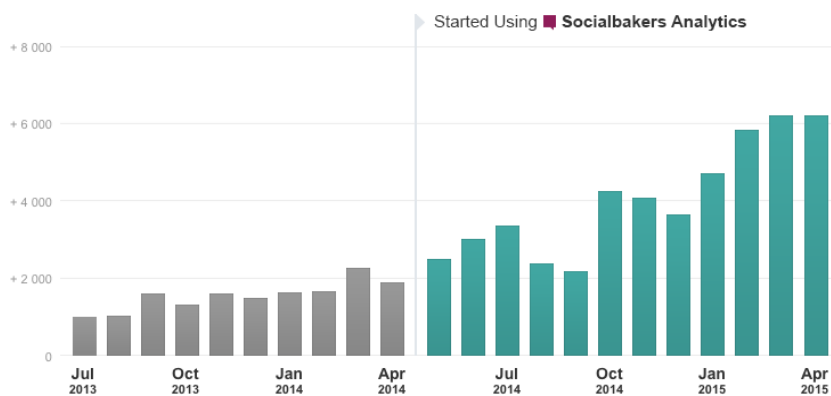

*Figure 2. Kenya's follower growth on Twitter[8]*

The Umati[9] project emerged out of recognition that mobile and digital technologies played a catalyzing role in Kenya's 2007/08 post-election violence. This project seeks to better understand the use of dangerous speech in the Kenyan online space. The project monitors particular blogs, forums, online newspapers, Facebook and Twitter. Online content monitored includes tweets, status updates and comments, posts, and blog entries. Set up in September 2012 ahead of Kenya's 2013 general elections, the Umati project sought to identify the use and role of social media in propagating dangerous hate speech online, something that at the time no other monitoring group was looking into. In order to understand the changes in online inflammatory speech used over time, the Umati project developed a contextualized methodology for identifying, collecting, and categorizing inflammatory speech in the Kenyan online space.

The media's near-unlimited growth, rapidly changing content, linguistic complexity, and often implicit context presents considerable challenges for data analytics. Meaningful analysis of the data can be achieved by collaboration of social science, computer science, especially Machine Learning and Text Data Mining, and quantitative analysis. The current work demonstrates how state-of-the-art statistical topic modelling methods can be used in analysis of Twitter data. Topic models are mostly unsupervised, data-driven means of capturing main discussions happening in collections of texts. Each topic is represented through a probability distribution over words occurring in the collection such that words that co-occur frequently are each assigned high probability in a given topic (Aletras, Baldwin, Lau, Stevenson, 2015). Statistical topic modelling is commonly based on automated generation of terms and term phrases associated with a given topic (Lau, Newman, Karimi, & Baldwin, 2010). It has been shown that terms provide for more relevant document retrieval than term phrases (Aletras, Baldwin, Lau, Stevenson, 2015).

We apply Latent Dirichlet Allocation (LDA) (Blei, Ng,, Jordan, 2003). To evaluate the LDA results we applied Normalized Mutual Information (NMI) and topic coherence analysis. The two measures are complementary by design: NMI rates representativeness of the constructed topics in the data set, whereas topic coherence

---

[8] http://www.socialbakers.com/resources/client-stories/airtel-kenya/

[9] KiSwahili word that means "crowd".

measures each topic individually.

We worked with four sets of Twitter texts provided by the Umati team[10]. Data collection was done utilising Twitter's Streaming API. Each data set represents a collective response to a significant social event in early 2014 (e.g., assassination of a controversial cleric, explosions in a major market, and attacks in two different communities).

## 2. Data Sets

We worked with four sets of texts collected from Twitter, each of which had generated online controversy but with varying sample sizes. The data were collected by the Umati project[11] through Twitter's Streaming API:

1. The Gikomba Twitter data mainly covers a bombing incident in a Nairobi market called Gikomba. The 482 tweets talk about explosions, blasts (i.e. what happened in the market), speculate on suspected perpetrators, and discuss how people feel about the bombing incident and how social organizations and the government reacted and responded to the incident.

2. The Mandera Twitter data contains tweets mainly talking about so-called "tribal clashes" in Mandera region, a Kenyan town located near the border with Somalia and Ethiopia. The data has 915 tweets in total.

3. The Makaburi dataset contains 20462 tweets. In those tweets, people are talking about the violent death of Sheikh Makaburi, a controversial Muslim preacher. As a result of the controversy of the late cleric and his public assassination, two opposite opinions exist towards the incident: some people think that killing is justified, while others think Makaburi was innocent.

4. The Mpeketoni data set is the largest, with 106348 tweets. In that data, people mainly discuss an attack that happened in Mpeketoni town in the coastal region of Kenya, as well as social and political problems behind that attack.

In the present study, we worked with tweets written in English (tweets written in other languages were left for future work). Our choice is not unreasonable. We know from other studies that automated language identification showed 81% of tweets in Kenya were in English[12]. Only 5% tweets were in Kiswahili. The rest were in an array of other languages including Hindi, Kikuyu, Somali, Luo, the Sheng dialect, and other languages. Many of these were mixed with English[13]. The Twitter data is considered to be more challenging than other social data due to a character limit and misspellings, slang, and informal capitalization (Eisenstein, 2013). All the four sets exhibit misspelling, slang, a large number of URLs, informal capitalization; typical for Twitter data. Those characteristics can significantly diminish capacity of the text analytics tools. Thus a considerable normalization effort is required to prepare the data for meaningful topic modelling. We performed data pre-processing with the aim to make the text suitable for application of automated Natural Language Processing techniques. The first step was to identify and delete duplicate text. Since this part of the study aims to analyze English text, on the next step we identified and filtered out messages in other languages. Finally, we deleted URLs, stop words (the, and, of), short words < 3 characters and tokenized each sentence.

---

[10] www.ihub.co.ke/umati
[11] http://ihub.co.ke/research/projects/23
[12] http://www.economist.com/blogs/baobab/2014/05/twitter-kenya
[13] http://whiteafrican.com/2014/05/01/what-twitter-can-tell-us-about-african-cities/

Traditional syntactic and semantic text analysis methods, i.e., technology and science known as Natural Language Processing (NLP), were developed to process contrived, well-edited text. Twitter messages, however, are often spontaneously written and unaltered after posting. To exemplify challenges faced by traditional NLP tools in analysis of Twitter data, we report Part-of-Speech (POS) tagging results by LingPipe, a commonly used NLP toolkit[14]. In many cases, the assigned POS tags are questionable and cannot be relied upon in further analysis of the data, e.g. "mall" was incorrectly identified as an adjective.

In further steps of data analysis, we employ Latent Dirichlet Allocation (LDA), statistical topic modelling technique (Blei, Ng, & Jordan, 2003). This unsupervised algorithm has been proved to be an effective probabilistic model for topic modelling of Twitter data.

# 3. Empirical Analysis

## 3.1 LDA application

Latent Dirichlet Allocation (LDA) is a topic modelling algorithm used for extracting topics from a given collection of documents (Blei, Ng,, Jordan, 2003). It builds models in unsupervised mode, i.e., does not need labelled training data. Based on the assumption that a document contains a mixture of N underlying different topics, and the document is generated by these topics with different proportions or probabilities, LDA is able to find out the topics and their relative proportions, which are distributed as a Latent Dirichlet random variable. The algorithm's performance can be managed though assumptions on the word and topic distributions. Table 1 reports on the adjustable parameters of the LDA model:

**Table 1: LDA adjustable parameters.**

| Variables | Definition |
|---|---|
| N | Required topic number |
| A | Dirichlet prior on the per-document topic distributions |
| B | Dirichlet prior on the per-topic word distributions |
| minTokenCount | Minimum token count (i.e. if one word appears less than minTokenCount, ignore that word in the document) |
| numTokens | Number of tokens of a data |
| numDocs | number of documents (in our experiment, which is number of records of a data) |

Given a pre-set of parameters such as α, which is the parameter of the Dirichlet prior on the per-document topic distributions and β, which is the parameter of the Dirichlet prior on the per-topic word distribution, the only required input is the documents and fixed topic number N. The output of LDA contains two parts. Part one is N topics with a list of words and count numbers for each word. Part two, for each document, indicates which topics it might belong to and the relative probability. The general processing procedure using LDA is as follows:

1. Setting parameters. The default parameters are $\alpha = 0.1$; $\beta = 0.01$; minTokenCount = 5;
2. Choose a topic number N and input data file.
3. Tokenize for the text data.
4. Run LDA algorithm to get the latent structure behind the text and print out the result.

---

[14] LingPipe Website: http://alias-i.com/lingpipe/

For the Gikomba data, we started with default parameters $\alpha = 0.1$; $\beta = 0.01$ and input parameter topic number N = 10, 15, 20 which means 10, 15, 20 desired topics. By comparing the Gikomba LDA result above, we choose topic number N = 15 as a basic group for further comparison since when N = 15, most topics have enough words to reveal information about the topic while without too much words to make the topics messy. In the next step of our experiment, we set N = 15 and tuning parameter $\alpha$ and $\beta$ by setting $\alpha = 0.1, 0.05, 0.2$ while $\beta = 0.01, 0.015, 0.007$ to see if the results show any difference.

For the Mandera data, we also start with default parameters $\alpha = 0.1$; $\beta = 0.01$ and topic number N = 10, 15, 20. Also we continue the experiment with tuning parameter $\alpha = 0.1, 0.05, 0.2$ and $\beta = 0.01, 0.015, 0.007$. Mandera LDA results with topic number N = 15 and 20 both show good results. However for the larger datasets, i.e. Makaburi and Mpeketoni data, fifteen topic numbers is not enough to capture sufficient information about the data. As a result, we tried topic number N = 20, 25 and 25, 30, 35, 45, 50 for Makaburi and Mpeketoni data respectively.

Since we have tried LDA algorithm with different parameters, in the following analysis, we use LDA (topics, $\alpha$, $\beta$) to denote LDA with different topic number, $\alpha$ (Dirichlet prior on the per-document topic distributions) and $\beta$ (Dirichlet prior on the per-topic word distributions). For example, LDA (15, 0.1, 0.01) means that we applied LDA with 15 topics, $\alpha = 0.1$ and $\beta = 0.01$.

## 3.2 Manual analysis of the LDA topics

From the manual analysis of the extracted topics, we confirmed that the LDA results are able to identify the event and reveal some relevant information about the said event. For example, in Gikomba LDA (15, 0.1, 0.01), topic 0: "blast", "bomb" give us an idea of what happened there, "kenya" indicates the location of the incident. In topic 1, the outliers are "security", "ban" (background: ban refers to restricting access to the place in question). In topic 2, "terror", "gikomba", "Nairobi", "market" indicate the location details. In topic 4 and topic 13, there are "clothes" and in topic 11 there are "nguo", which is also "clothes" in Swahili; both words show how the bomb was hidden, that is, in clothes (background: the section of Gikomba market that was attacked primarily deals with open-air trading of clothes).

Although we considered the pre-processed Gikomba data to be our benchmark (Section 2), we also applied LDA on other versions of the Gikomba data: one dataset keeps short words while removing stop words; another dataset keeps stop words and short words. Comparing the obtained LDA results, the found commonalities are as follows:
1. All results reveal background information about the incident, including the location (by words "Gikomba", "Nairobi", "Market" etc.), the nature of the incident (by words

"fire", "bomb", "attack", "blast" etc.), actors in the attack, and how the attack happened (by words "suspect", "terror", "clothes/nguo" etc.).
2. All results contain a topic with word "security" which means people care and talking about security issues after the attack.
3. All results contain topic with references to people's names, such as "Kimaiyo" (inspector general of the Kenya police), "robertalai" (a prominent investigative blogger), "kidero" (governor of Nairobi), etc.

At the same time, there exist notable differences between the results:
1. When the data contains short words, "security" often appears in a topic, which also lists "no"; we assume this joint occurrence indicates the phase "no security". As expected, in the data with short words filtered out, the word "no" disappeared from the resulting topics.
2. Some high frequency words appear in LDA results obtained on one data, but not in others. For example, according to our preliminary words' frequency analysis, "reconstruction" and "donate" have high frequency. However, they only appear in the LDA results obtained on the data version containing stop words and short words; both words appear together with word "kidero" (background: Evans Kidero, the Nairobi County Governor donated money to reconstruct Gikomba market after the attack). In the LDA results obtained on the other two data versions, only "kidero" appears.

3. The LDA results on the benchmark data (i.e., stop and short words removed) contain more references to politicians and government officials, such as "kimaiyo", "robertalai" (a prominent blogger Robert Alia, who was often the first to "break" stories on Twitter) and "railaodinga"; these references are absent in the other results.

Regarding the Mandera LDA results, outliers in the topics are shown as following:
In Mandera LDA (15, 0.1, 0.01), topic 0, "fighting", "somalis" as well as in topic 1, "clashes", "tribal", "killing" indicate the situation in Mandera (background: there exists serious conflict between different clans in Mandera). In topic 3, we found that a word "sad" appears together with "killing". This joint occurrence may indicate people's feeling towards the conflict. Topic 4 lists "security", "insecurity", "need", "government", "violence", "end". This implies that people hope for government to do something to ensure their security and end the conflict situation. Topic 5 lists "muslim" which indicates religion as an important factor in the discussion. Topic 8 and topic 12 list "somalis". We found it interesting that Somalis and Muslim never appear in the same topic in this dataset. Perhaps, tweets do not mention them together when discussing events in Mandera. In topic 14, we assume from words "solution", "peace", "leaders" that the topic is about ways to achieve peace. The extracted topics also contain references to people, e.g. "johnallannamu" (a renowned journalist), "uhuru" (president of Kenya), "duale" (a prominent politician) and "AbdulazizHOA" (an Al Shabaab terrorist group follower).

Comparing the benchmark Mandera LDA results (i.e., stop words and short words are deleted) with the results obtained on other data versions, we found that all LDA results can reveal the situation in Mandera. All the results contain description of clashes, killings and violence. At the same time, the results contain topics referring to peace and finding solutions for the situation. The most noteworthy difference between the results is that, sentiment-bearing words such as "great", "sad" appear only when the stop words are deleted from original Mandera data.

For Makaburi data, a much larger dataset, we chose Makaburi LDA(25, 0.1,0.01) as the baseline result. In Makaburi LDA (25, 0.1,0.01), words in topics also reveal information about what happened to Sheikh Makaburi. From topic 0, "death", "gun" and in topic 8, "shoot", we understand that Makaburi was fatally shot. Some topics also reveal background of Makaburi's assassination for example in topic 0, "religion", "Islam" and in topic 12 "Christians", "Muslims" indicate there might be some religious issues that have connections with Makaburi's death. Remarkably, different topics contain words with opposite sentiments ("deserved" vs "sad"). We can assume that when people talked about Makaburi's death, they may have had antagonistic opinions towards him. Some antonyms even appear in the same topic. For example in topic 3, we have "good" and "bad". In topic 24, there are "heaven" and "hell". Other examples including in topic 0, there is "deserved" while in topic 4, there is "innocent". In topic 15, there is "happy" while in topic 22 there is "sad" etc. Besides that, we can see people's concern about security via topic 2, which lists "security", "situation" and "government". When we reduced the number of topics to 15 and built Makaburi LDA (15,0.1,0.01) we found that almost every topic contains the word "makaburi". As in previous results, the extracted topics contain opposite sentiment words pairs "good" and "bad", "heaven" and "hell", "innocent" and "deserve".

Due to the large volume of tweets in the Mpeketoni dataset, we chose Mpeketoni LDA (30,0.1,0.01) as a baseline result and found it shows more clear topics than the other datasets, i.e., when looking through a topic, it is easier for us to assume what the topic means. For example, in topic 0, since it contains "mpeketoniattack", "lost", "families", "affected", "sad", "prayers", "peace", we can conclude that in the topic, people are mainly talking about the victims or families affected by the attack, and express their condolences as well as their wish for peace. Topic 2 is about security issues since it contains "safe" and "security" but it may be that people are discussing more about solutions or policies rather than just expressing a wish for security since we found "president", "never", "uhuru" and "need". In topic 4, there are "mpeketoniattack", "god", "country", "pray", "help", "peace", "sad" from which we can feel people's emotion towards the attack and their call for help. In topic 8, people are talking about government's reaction to the attack since we have "police", "response", "officers", "military" and "deployed". Topic 9 shows that people are unsatisfied with government since in the topic, the word "blame" has high frequency and also there are "politics", "insecurity", "government", "need" and "instead". Similarly, in topic 10, which contains "government", "security", "citizens", "protect", "innocent", "failed" and "die" and "sad", the high frequency of these words likely indicates a strong expectation for the government, as well as sadness for some failure and death. In the Mpeketoni data, people also talking about politicians but unlike in other datasets, names of those politicians tend to appear in the same topic. One example is topic 12, which contains "ukenyatta" (president Uhuru Kenyatta), "joelenku" (Cabinet secretary of the interior Joseph Ole Lenku), "railaodinga" (opposition politician Raila Odinga) and "williamsruto"(deputy president William Samoei Ruto).

We can assume that topic 15 identifies suspects in the attack since "shabaab", "alshabaab", "responsibility", "claim" are the highest frequency words in that topic. Topic 19 focuses on media since there are "ktnkenya", "citizentvkenya", "ntvkenya", "media" and "citizentvnews" appear together. Topic 22 reveals information to describe the attack incident in Mpeketoni via words such as "lamu", "police", "station", "hotels", "gunmen" and "fire" (background: gunmen attacked hotels and a police station in

Mpeketoni near Lamu town, and set fire to several buildings). Topic 24 talks about conflict between different communities since it is represented through "killed", "kikuyu", "ethnic", "tribe" and "community". Another interesting topic is topic 25, which contains "consulate", "british", "closed", "west", "info" and "know". (Mpeketoni attacks happened soon after the British consulate at the coastal city of Mombasa was closed; as a result people may discuss whether this happened by chance or was pre-meditated.)

The number of topics is critical to the quality of analysis. When we reduced the number of topics to 15, Mpeketoni LDA (15,0.1,0.01) did not extract as well-identified topics as Mpeketoni LDA (30,0.1,0.01) had done. When we increased the number of topics to 50, the model Mpeketoni LDA (50,0.1,0.01) faulted to improve the quality of the topic extraction; it replicated many of the topics found in the output of the models with a smaller number of topics. For example, topic 2 of Mpeketoni LDA (50,0.1,0.01) is similar to topic 25 of Mpeketoni LDA (30,0.1,0.01). In that topic, people speculate on the coincidence of the Mpeketoni attacks being preceded by the closure of the British consulate in Mombasa city, citing security concerns. Topic 16 of Mpeketoni LDA (50,0.1,0.01) is similar to topic 0 of Mpeketoni LDA (30,0.1,0.01); in that topic, people express their condolences and sadness for the lives lost in the attacks. In Mpeketoni LDA (50,0.1,0.01), there also exists some topics showcasing similar information. For example, topic 10 is similar as topic 16; topic 0 is similar to topic 8.

## 3.3 Normalized Mutual Information results

In the experiment we used NMI (Normalized Mutual Information) to evaluate overall documents (tweets) cluster quality. The following formula is used to calculate NMI (Mehrotra et al, 2013),:

$$NMI(X,Y) = \frac{2I(X;Y)}{H(X)+H(Y)}$$

where $I(X;Y)$ is mutual information between X and Y, where $X = \{X1, X2, ...Xn\}$ and $Y = \{Y1, Y2,...Yn\}$. Xi is the set of tweets in LDA's topic i while Yj is the set of tweets with the label j. In our experiments, a tweet with the label j means that tweet has the highest probability of belonging to topic j; n is the number of topics. $I(X;Y)$ is

$$I(X;Y) = \sum_{y \in Y} \sum_{x \in X} p(x,y) \log(\frac{p(x,y)}{p(x)p(y)})$$

In the formula, $p(x_i)$ means probability of being classified to topic i, $p(y_j)$ means probability of labeled to topic j while $p(x_i,y_j)$ means probability of being classified to cluster i but actually labeled to cluster j. $H(X)$ is entropy of X as calculated by the following formula:

$$H(X) = -\sum_{i=1}^{n} p(x_i) \log_2 p(x_i)$$

NMI = 0 means the clustering result is totally different from the label, while NMI = 1 means clustering result and label result are identical. In our experiments, we evaluated NMI of LDA with different topic numbers. Table 2 reports the results:

**Table 2: NMI results for LDA models.**

| LDA models | NMI results |
|---|---|
| Gikomba(10,0.1,0.01) | 0.606 |
| Gikomba(15,0.1,0.01) | 0.540 |
| Gikomba(20,0.1,0.01) | 0.540 |
| Mandera(10,0.1,0.01) | 0.535 |
| Mandera(15,0.1,0.01) | 0.488 |
| Mandera(25,0.1,0.01) | 0.463 |
| Makaburi(5,0.1,0.01) | 0.563 |
| Makaburi(15,0.1,0.01) | 0.406 |
| Makaburi(25,0.1,0.01) | 0.364 |
| Mpeketoni(15,0.1,0.01) | 0.396 |
| Mpeketoni(30,0.1,0.01) | 0.331 |
| Mpeketoni(50,0.1,0.01) | 0.290 |

The results show that with fewer topics, the NMI value tends to be higher. Since NMI presents similarity of clustered tweets set and labelled tweets set, the overall NMI results indicate that with fewer topics, tweets set are more correctly clustered. The reason for this phenomenon could be the length of each document (tweet) is much shorter if compared to traditional documents. Since the length for each tweet is limited (usually no longer than 140 characters), information contained in a single tweet is also limited. Hence, when the number of topics increases, many topics tend to contain the same words; as a result, it is hard to determine to which topic a document be assigned. In further experiments, we can use different tweeter pooling schemes (Mehrotra, Sanner, Buntine, Xie, 2013) and see whether they affect the NMI results.

## 3.4 Topic coherence analysis

Topic coherence measures each topic by scoring it based on calculating the degree of semantic similarity between words in the topic. It is often considered as a metric to evaluate the quality of a topic.

In our experiment, we use the following formula (Nugroho, Molla-Aliod, Yang, Zhong, Paris and Nepal, 2015) to implement topic coherence evaluation:

$$Co(k,W) = \sum_{m=2}^{M}\sum_{l=1}^{m-1} \log \frac{T(w_m, w_l)+1}{T(w_l)}$$

In the formula, k and W mean that in topic k, the total words set is W. $T(w_m,w_l)$ is the number of documents containing both word m and word l while $T(w_l)$ is the number of documents containing word l. Topics with higher values are considered more coherent, thus are better.

Since we have tried LDA on different parameters including different α, β with default topic number and different topic numbers with default α, β, we then evaluate those results by calculating the normalized average topic coherence value (average coherence value between each topic), as well as standard deviation between normalized coherence value for each topic (average coherence value between each word in each topic).

We use the following formula, which evolves from the previous topic coherence formula to calculate normalized average topic coherence value:

$$Co1(k, W) = (\sum_{m=2}^{M} \sum_{l=1}^{m-1} \log \frac{T(w_m, w_l) + 1}{T(w_l)}) / n$$

In the formula, n is number of words in topic k.

The following formula is used to calculate standard deviation:

$$SD = \sqrt{\left| \frac{\sum_{k=1}^{N} Co_k^2}{N} - \left(\frac{\sum_{k=1}^{N} Co_k}{N}\right)^2 \right|}$$

In the formula, N refers to topic number, while $Co_k$ is the topic coherence of topic k. Topic coherence for the Gikomba and Mandera datasets is reported in the following Tables 3 and 4. The reported results show that LDA with α =0.05 provides for a slightly better coherence than other α, but still does not make a significant difference. We also varied the number of topics in the built models. We report assessment of the LDA results with different topic numbers in Table 5.

**Table 3: Topic coherence for Gikomba data with different α and β**

| Parameters | Co |
|---|---|
| Gikomba LDA (15, 0.05,0.007) | 0.895 ± 0.298 |
| Gikomba LDA (15, 0.05,0.01) | 0.933 ± 0.355 |
| Gikomba LDA (15, 0.05,0.015) | 0.906 ± 0.255 |
| Gikomba LDA (15, 0.1,0.007) | 0.751 ± 0.313 |
| Gikomba LDA (15, 0.1,0.01) | 0.765 ± 0.348 |
| Gikomba LDA (15, 0.1,0.015) | 0.811 ± 0.344 |
| Gikomba LDA (15, 0.2,0.007) | 0.552 ± 0.318 |
| Gikomba LDA (15, 0.2,0.01) | 0.52 ± 0.284 |
| Gikomba LDA (15, 0.2,0.015) | 0.628 ± 0.427 |

**Table 4: Topic coherence for Mandera data with different α and β**

| Parameters | Co |
|---|---|
| Mandera LDA (15, 0.05,0.007) | 1.571 ± 0.542 |
| Mandera LDA (15, 0.05,0.01) | 1.54 ± 0.457 |
| Mandera LDA (15, 0.05,0.015) | 1.47 ± 0.517 |
| Mandera LDA (15, 0.1,0.007) | 1.256 ± 0.477 |
| Mandera LDA (15, 0.1,0.01) | 1.292 ± 0.434 |
| Mandera LDA (15, 0.1,0.015) | 1.318 ± 0.379 |
| Mandera LDA (15, 0.2,0.007) | 1.096 ± 0.373 |
| Mandera LDA (15, 0.2,0.01) | 1.049 ± 0.232 |
| Mandera LDA (15, 0.2,0.015) | 1.061 ± 0.464 |

**Table 5: Topic coherence for the four data sets with different topic numbers**

| Parameters | $Co$ |
|---|---|
| Gikomba LDA (10, 0.1, 0.01) | $1.126 \pm 0.412$ |
| Gikomba LDA (15, 0.1, 0.01) | $0.765 \pm 0.412$ |
| Gikomba LDA (20, 0.1, 0.01) | $0.562 \pm 0.412$ |
| Mandera LDA (10, 0.1, 0.01) | $1.739 \pm 0.426$ |
| Mandera LDA (15, 0.1, 0.01) | $1.292 \pm 0.434$ |
| Mandera LDA (25, 0.1, 0.01) | $0.907 \pm 0.375$ |
| Makaburi LDA (5, 0.1, 0.01) | $3.782 \pm 0.6$ |
| Makaburi LDA (15, 0.1, 0.01) | $2.798 \pm 0.724$ |
| Makaburi LDA (25, 0.1, 0.01) | $2.375 \pm 0.673$ |
| Mpeketoni LDA (15, 0.1, 0.01) | $3.976 \pm 0.696$ |
| Mpeketoni LDA (30, 0.1, 0.01) | $3.39 \pm 0.745$ |
| Mpeketoni LDA (50, 0.1, 0.01) | $2.883 \pm 0.747$ |

Manually assessing the results with different topic numbers, we found that Gikomba LDA (15, 0.1, 0.01) provides for better results. Most topics contain enough words to reveal information and not too many to make the topic unreadable. For Mandera data, Mandera LDA (15, 0.1, 0.01) also outputs comprehensible results. Hence, choosing 15 topics worked well for small data sets. Makaburi and Mpeketoni LDA models yield different results. Since data volumes are considerably larger for both datasets, increase in the topic number improves the modelling results. Makaburi LDA (25, 0.1, 0.01), Mpeketoni LDA (30, 0.1, 0.01) and Mpeketoni LDA (50, 0.1, 0.01) all output reasonably understandable topics.

In topic coherence analysis, however, the lower topic number in Mandera LDA (10, 0.1, 0.01) seems to have better coherence performance than Mandera LDA (15, 0.1, 0.01). For Makaburi and Mpeketoni data, coherence performance also seems better with a smaller topic number. By examining topic coherence values for each topic and the appearance count for each word in the topic, we found that for fewer topic numbers, each topic tends to have a higher score and words in each topic have larger appearance count numbers thus causing a higher total coherence value. This can be explained as for fewer topic numbers, for each document, probability of belonging to a particular topic increases. Hence each topic has more words assigned to it and those words have a higher chance to appear in a document assigned to the same topic.

When examining topics with higher coherence values, we found those high value topics usually contain high frequency words. Those words often reveal the event's essential information. For example, in Gikomba LDA (15, 0.1, 0.01), the highest topic coherence value is 2.162 for topic 8. In topic 8, words with the greatest contribution to its coherence value are "Gikomba" and "market". However, in outlier topics, for example, in topic 3, the word "security" reveals what people care about but the word only co-occurs once with "Gikomba", "attack" and "business".

Based on this and similar examples, we conclude that LDA is able to identify rare topics, i.e., topics that exhibit a well-defined content while appearing infrequently in the data. We consider this ability to be a significant advantage of the LDA algorithm.

# 4. Related Work

Unsupervised topic modelling can be defined as a search for patterns in the textual data. Patterns can be difficult to be identified a priori, thus, making assessment of the built models challenging. Using manual evaluation can introduce a human bias, hence it is imperative to supplement it with evaluation measures (Sokolova and Lapalme, 2009). When patterns do not need to have a semantic interpretation, one evaluation approach is to fix a number of patterns and then separate different patterns (Tuytelaars, Lampert, Blaschko, Buntine, 2010.) However, many studies (e.g., textual data, social media studies) expect semantic interpretation of the found topics. In those cases, we can evaluate topic modelling by calculating topic coherence that scores a single topic by measuring the degree of semantic similarity between words in the topic.

Several studies compare topic coherence of different topic modelling algorithms. Considerable research has been done to compare LDA modelling quality with other methods, e.g. Non-negative Matrix Factorization (NMF) (Lee, Seung; 2001). Consider (Nugroho, Molla-Aliod, Yang, Zhong, Paris and Nepal, 2015); in this work, after the topic extraction, the authors calculate coherence for each topic, which we also use in our experiment. Besides that, they also evaluates tweet-topic accuracy by using the result of manually labelled training dataset to compare with the automated classified tweet and using F-score to compute the harmonic mean of both precision P and recall R. Their empirical results show that LDA has better topic coherence performance over NMF (LDA has topic coherence values ranging from 38.39-58.39, while NMF results range from 37.82-54.04). However both LDA and NMF are not comparable with intLDA (NMI range from 41.27-59.12), which the authors advocate in the article as an improved LDA algorithm.

UCI and UMASS measures are used to evaluate topic coherence in Stevens, Kegelmeyer, Andrzejewski, and Buttler (2012). For UCI measure, which indicates average coherence score, the result shows LDA and NMF almost the same and stable at around -1.5. However as for entropy for the UMASS score, NMF produces unstable results ranging from 0.3 to 5, i.e. NMF learns topics with different levels of quality, some with high and some with very low quality. Besides NMI, UCI and UMASS, point-wise mutual information (PMI) is also used to evaluate topic coherence (Mehrotra, Sanner, Buntine, Xie, 2013). Work by Chen and Liu (2014) aims at addressing the topic coherence issue existing in unsupervised models. The authors propose an automatic process to learn prior knowledge from various domains and use that knowledge to generate more coherence topics. Further, the authors propose a new knowledge-based topic model LTM to deal with possible incorrect knowledge.

Quality of Twitter's topic modelling can be improved through various approaches. A commonly used method is pre-processing data through tweet-pooling schemes. For example in the work by Mehrotra, Sanner, Buntine, and Xie (2013), the authors provide an automatic hashtag assignment scheme to improve LDA topic quality, which proved in the later experiment to be the best pooling scheme with PMI value increased from 0.47 to 1.07. However in our Twitter data sets, most tweets do not have a hashtag so it is infeasible for our experiment to verify the effect of hashtag pooling. In other work, Nugroho, Molla-Aliod, Yang, Zhong, Paris and Nepal (2015) propose intLDA as a variant of LDA, incorporating the tweet relationship to improve the tweet-topic distributions.

In the work by Xie, Yang, and Xing(2015), the authors build a Markov Random Field (MRF) regularized LDA model, which defines a MRF on the latent topic layer of LDA in order to create better chance for similar words to appear in the same topic. Besides LDA, there also some works targeting at other topic modeling methods such as the work by Yan, Guo, Liu, Cheng, and Wang (2013), in which the authors improve NMF algorithm by directly estimating topics from term correlation data rather than the sparse term-document matrix.

Twitter has become one of the major research sources in text mining field over the past years; it differs from traditional media data sources due to large volumes and the short length of each document (tweets). The work by Zhao and Jiang (2011) provides comprehensive comparison between Twitter and traditional news media content analysis through LDA topic modelling. Eisenstein (2013) focuses on one of the most typical attributes of online data: the high abundance of misspellings, abbreviated phrases, and Internet slang or shorthand. The author analyzes different types of "bad language" and their possible causes, and then provides suggestions on how to mitigate it, such as normalization and preprocessing. Twitter data has been used in some business applications. For example Si, Mukherjee, Liu, Li, Li, and Deng (2013) use topic-based sentiments from Twitter to help predict the stock market. The authors first utilize a continuous Dirichlet Process Mixture model to learn the daily topic set and then build a sentiment time series base on the opinion words distribution of each topic. Finally, the authors use the stock index and the Twitter sentiment time series to predict the market.

# 5. Conclusions and Future Work

With a large number of people participating on social networks and online forums, social media has become a major source for social data mining. In the presented work, we applied statistical topic modelling to extract and analyze content of Twitter data.

We worked with four sets of Twitter messages collected in Kenya for the Umati project (monitoring online dangerous speech). Each dataset follows a specific event. For the topic modelling, we applied Latent Dirichlet Allocation. We varied the LDA parameters to find a model that outputs more informative and coherent topics (as evaluated by NMI and topic coherence analysis). Performance of the LDA models was not affected by changes in distribution parameters $\alpha$ and $\beta$. At the same time, the results significantly changed with the change of topic numbers. As we expected, the quality of LDA results also depends on the amount of records in the data.

Manual analysis of the results revealed that LDA is able to extract detailed information from the data. It extracts all the major event components, including the geographic location, people involved, how the event unfolded etc. It is also important to note that all the extracted topics were related to the events covered by the collected data. Our method does not confide to the analysis of Kenya's data. It can be applied to the analysis of Twitter data collected in similar settings . Understanding the influence of social networks can help political scientists to better understand how such information can be used not only in the dissemination of online violent messaging, but can also help to inform responses that could help to mitigate or prevent violence from starting and escalating. It can also act as an early warning mechanism for when incidents do occur.

With a more real time understanding of emerging issues, programs that aims to prevent and respond to conflict escalation can consider how to respond to current and emerging threats. Relevant questions would include:   Why and when do online tools get used to incite violence and hate?  How does the dissemination of information about violence in online spaces affect communities at-risk? What is the nature of online speech in these contexts? Is it different than offline content in terms of its impact, and if so, then how? How can we devise and improve upon early warning mechanisms to prevent the escalation of conflict? How can online tools and narratives be instead used to respond to violence constructive and promote understanding?

The answers to these questions can help civil society, citizens and government to create more constructive policies and programs to help reduce the risk of violence. Organizations like Radio La Benevolencija which helped Rwandans to understand how hate speech enabled the genocide through radio soap operas can consider both how online spaces affect these risks and how they can be used to complement existing strategies. Policy makers can consider responsible policies to monitor and regulate hate speech and violent speech in online spaces, without infringing on civil rights. Community leaders in online and offline spaces can create better strategies for responding to and leveraging social media tools, narratives and strategies.

In future, the goal of improved text conceptualization can be achieved through improvement of the automated methods.  Topic modelling by MRF (Markov Random Field) -regularized LDA (Xie, Yang, Xing, 2015) which defines a MRF on the latent topic layer of LDA can create a better chance for similar words to appear in the same topic. We can also apply Non-Negative Matrix Factorization and compare the resulting topics with LDA's. Also, NMI and topic coherence together would help improve the quality of topic modelling, since they enable us to evaluate different model from both overall and topic-specific point of view. For example, when comparing NMF and regular LDA, we can input the same topic number and try to find out which model get higher NMI and topic coherence values. When comparing regular LDA and improved LDA, we can also apply NMI and topic coherence under the same LDA parameters (topic number, $\alpha$, $\beta$).

# References

Aletras, N., Baldwin, T., Lau, J., Stevenson, M., Evaluating Topic Representations for Exploring Document Collections, *Journal of the Association for Information Science and Technology*, 2015

Blei, D.M., Ng, A.Y., & Jordan, M.I. Latent Dirichlet Allocation. *Journal of Machine Learning Research,* 3, 993–1022, 2003.

Chen, Z., B. Liu, "Topic Modeling using Topics from Many Domains, Lifelong Learning and Big Data", *Proceedings of the 31st International Conference on Machine Learning (ICML-14).* 2014

Eisenstein, J. "What to do about bad language on the Internet", *HLT-NAACL*, 2013

Lau, J.H., Newman, D., Karimi, S., & Baldwin, T. (2010). Best topic word selection for topic labelling. *In Proceedings of the 23rd International Conference on Computational Linguistics (COLING 10) (pp.* 605–613*).*

Lee, D., and H. Sebastian Seung. "Algorithms for non-negative matrix factorization." *Advances in neural information processing systems*. 2001.

Matwin, S. (2013) Institute for Big Data Analytics: Message from the Director, https://bigdata.cs.dal.ca/about, retrieved July 28, 2016.

Mehrotra, R., S. Sanner, W. Buntine, L. Xie, "Improving LDA Topic Models for Microblogs via Tweet Pooling and Automatic Labeling", *Proceedings of the 36th international ACM SIGIR conference on Research and development in information retrieval*. 2013

Nugroho, R., D. Molla-Aliod, J. Yang, Y. Zhong, C. Paris and S. Nepal, "Incorporating Tweet Relationships into Topic Derivation", *Proceedings of the 2015 Conference of the Pacific Association for Computational Linguistics, PACLING*. 2015

Ramisch, J. J., Misiko, M. T., Ekise, I. E., & Mukalama, J. B. Strengthening 'folk ecology': community-based learning for integrated soil fertility management, western Kenya. *International journal of agricultural sustainability*, *4*(2), 154-168, Taylor & Francis, 2006

Si, J., A. Mukherjee, B. Liu, Q. Li, H. Li, X. Deng, "Exploiting Topic based Twitter Sentiment for Stock Prediction", *EMNLP* - 2014, pp. 1139-1145, 2014

Sokolova, M. and G. Lapalme, "A Systematic Analysis of Performance Measures for Classification Tasks", *Information Processing & Management*, 45, p. 427–437, 2009

Stevens, K., P. Kegelmeyer, D. Andrzejewski, D. Buttler, "Exploring Topic Coherence over many models and many topics", *Proceedings of the 2012 Joint Conference on Empirical Methods in Natural Language Processing and Computational Natural Language Learning* (pp. 952-961). 2012.

Tonkin, E., Pfeiffer, H. D. and Tourte, G. (2012), Twitter, information sharing and the London riots?. *Bul. Am. Soc. Info. Sci. Tech.*, 38: 49–57.

Tuytelaars, T., Lampert, C. H., Blaschko, M. B., & Buntine, W. (2010). Unsupervised object discovery, *International journal of computer vision*, *88*(2), 284-302.

Yan, X., J. Guo, S. Liu, X. Cheng, Y. Wang, "Learning Topics in Short Texts by Non-negative Matrix Factorization on Term Correlation Matrix", *Proceedings of the SIAM International Conference on Data Mining* , 2013

Xie, P., D. Yang, E. Xing, "Incorporating Word Correlation Knowledge into Topic Modeling", *Conference of the North American Chapter of the Association for Computational Linguistics*, 2015

Zhao, X., J. Jiang, "An Empirical Comparison of Topics in Twitter and Traditional Media", *Singapore Management University School of Information Systems Technical paper series. Retrieved November* 10 (2011) 2011